# Examples for counterintuitive behavior of the new citation-rank indicator P100 for bibliometric evaluations

**Michael Schreiber**

*Institute of Physics, Chemnitz University of Technology, 09107 Chemnitz, Germany.*
*Phone: +49 371 531 21910, Fax: +49 371 531 21919*
*E-mail: schreiber@physik.tu-chemnitz.de*

A new percentile-based rating scale P100 has recently been proposed to describe the citation impact in terms of the distribution of the unique citation values. Here I investigate P100 for 5 example datasets, two simple fictitious models and three larger empirical samples. Counterintuitive behavior is demonstrated in the model datasets, pointing to difficulties when the evolution with time of the indicator is analyzed or when different fields or publication years are compared. It is shown that similar problems can occur for the three larger datasets of empirical citation values. Further, it is observed that the performance evalution result in terms of percentiles can be influenced by selecting different journals for publication of a manuscript.

## 1. Introduction

Percentile-based indicators have become a standard tool for the evaluation of publications in terms of citation counts (Bornmann, Leydesdorff & Mutz, 2013). They are based on the position of a publication within a given reference distribution. For this purpose the papers in the reference distribution are usually sorted according to their (increasing) numbers of citations. It is then easy to determine the median, quartiles, deciles, or other percentiles, although there is an uncertainty which occurs in the application of different counting rules and an ambiguity which occurs when the calculated quantile value equals the threshold between the percentiles (Schreiber, 2013a). Applying the formula of Hazen (1914) does not solve this ambiguity (Schreiber, 2013b). Recently a new approach has been proposed which is based on the distribution of unique citation values (Bornmann, Leydesdorff & Wang, 2013). This means that the occurring citation frequencies in a dataset are sorted and then attributed to percentiles. Usually a large dataset contains many papers with the same number of citations which contribute thus only one value to the distribution for the new citation-rank indicator P100. Therefore a great reduction of the data is achieved (Bornmann & Mutz, 2014).

The advantages of the new approach are that one can easily attribute by definition the lowest scale value (0 for the P100 indicator) to the paper with the lowest citation count (usually zero citations) in a reference set and the highest scale value (100 in the P100 indicator) to the paper with the highest impact. Moreover, tied publications do not constitute a problem, because they have the same unique citation value and thus are attributed to the same percentile value by definition. The above mentioned ambiguity, however, remains: For example, if the percentile exactly coincides with the median then it is not clear whether the citation value belongs to the top 50% or the bottom 50%.

Bornmann and Mutz (2014) have already warned about reliability problems, because citation counts based on only one or two papers are not stable against random fluctuations. In my opinion, this is a severe difficulty, because such fluctuations are not necessarily random, but can easily occur in the time evolution of citation data. In particular in the high-citation-frequency tail of the distribution it is very likely that the unique citation counts are based on few papers and that many citation counts are not present at all in a given dataset. This will be demonstrated for three empirical examples, yielding counterintuitive behavior of P100 when small changes in the data base are implemented or when comparing different publication years. But before investigating these larger datasets, the effects are elicited for two small fictitious datasets.

Recently, Bornmann and Mutz (2014) have improved the P100 indicator by considering also the frequency of papers with the same citation counts in the ranking. However, this refinement changes the behavior drastically and therefore the resulting indicator P100' shall not be discussed here. A comparison between P100 and P100' for empirical examples will be the subject of a forthcoming study, which shall also include an analysis whether P100' is indeed an improvement in comparison to P100.

Another result of the present investigation is that one can improve the outcome of evaluations in terms of percentile-based indicators by choosing the proper journal for the publication of one's manuscript. This observation shall be substantiated in a forthcoming investigation, too.

## 2. The first fictitious model example: simulating time evolution

As a first example I consider a small reference set similar to the dataset which has been discussed by Bornmann and Mutz (2014). It comprises eight papers but only six unique citation counts, see Table 1. Accordingly there are 6 ranks $i = 0, 1, \ldots, 5$ attributed. The citation counts 0 and 5 are not ranked, because there are no papers with these citation counts; thus the respective unique citation counts are not comprised in the dataset, neither are the unique citation counts 6, 8, 9, 11, 12, ... relevant for the discussed datasets. The values of the P100 indicator corresponding to the remaining 6 unique citation counts are defined as $100 * i / i_{max}$ and also given in the table. Note that the ranks and the P100 values differ from those given by Bornmann and Mutz (2014), because I have excluded the uncited paper from their example. In this way one can better show the effect of their rule that the paper with the lowest citation count should receive the lowest scale value, i.e. one has rank 0 for the paper with the lowest impact, in this case the paper with 1 citation. If one now assumes that this paper with the lowest rank receives one more citation, only 5 unique citation values remain and the lowest rank 0 has to be attributed to the two papers with two citations. The results of this first modification are also given in Table 1 and one can see the surprising outcome, that beside the least cited and the most cited paper, the P100 indicator has decreased for all other papers in comparison with the original dataset. Assuming that the second lowest paper in the original dataset had received one citation less would mean a change in the opposite direction. It is interesting to note that this would have produced the same ranks and P100 values as given for the first modification in Table 1. I consider this to be a realistic example (except for the small total number of papers) because it simulates the possible time evolution of the dataset.

A second modification is also presented in Table 1 where one of the three papers with four citations in the original dataset is assumed to receive another citation. Consequently there are now seven unique citation counts as shown in Table 1 together with their P100 values. If we now speculate that the eight papers are authored by three scientists X, Y, Z as denoted in Table 1 then this second modification leaves the original P100 indicator for the papers of X unchanged, increases the result for Z, and decreases the original P100 values for all papers of Y except the one that has received the additional citation. While the outcome for X is as was to be expected, the increase for Z is surprising because it cannot be anticipated that an improvement for Y leads to a better value for Z. The decreases for Y are also unsatisfactory, because it seems that an improved impact leads overall to a worse evaluation of Y.

## 3. The second fictitious model example: comparing different fields

For the second example I have constructed a dataset with 100 papers and 11 unique citation counts, as shown in Table 2. The 100 papers are distributed over these citation counts in four different ways to simulate different fields. I have kept the number of singly cited papers constant at a value of 20. For increasing citation counts the number of papers in the first case A is approximately given by an inverse power law in terms of the number of citations. For the second case B I have halved these paper counts (and rounded to the higher integer). For the third case C I have increased the values from case A by 50% (rounded to the higher integer). For the fourth case D I have used the same paper counts as for case A, but shifted by one citation, i.e. one line downwards in Table 2. The number of uncited papers has been adjusted in all cases to yield a total number of 100 papers.

These rather drastic changes have no influence on the ranking in terms of unique citation counts because all citation frequencies from 0 to 10 are present in all four cases. Therefore there is also no change in the corresponding values of the P100 indicator. Consequently, although we have four different cases which might resemble four rather different research fields with rather different citation practices, this is not reflected by the P100 indicator. In my view this is rather unsatisfactory.

To be specific, let us look for the top 10% of the papers in terms of received citations. This is a widely used and important performance indicator in bibliometrics. For case A the threshold corresponds to 5 citations. Counting all papers at the threshold as belonging to the top decile, then all the 13 papers with 5 or more citations belong to the top 10%. For case B the threshold would already be reached with 4 citations, for case C only with 7 citations, and for case D with 6 citations; and 11, 10, 12 papers would be in the top-10% category, respectively. Thus the 4 examples are distinguished and these differences appear reasonable comparing the different citation distributions in the 4 cases. Similarly reasonable distinctions can be obtained for the top quartile of papers as well as for the median. In contrast, unfortunately the P100 indicator yields no distinction between the four cases.

A possible criticism to the example in table 2 could be, that it is not realistic that the most-cited papers in different fields have the same citation count. However, this criticism can be remedied by changing some or all of the four highest citation counts to different higher levels in the different fields. In this way the four different fields could be even more distinct and more realistically distinguished without changing the values of the P100 indicator. Due to the small paper counts for the four highest unique citation counts in Table 2, the restriction to keep the number of unique citation counts unchanged is not unreasonable. But as long as the number of unique citation counts is not changed, then the above discussion remains exactly the same. Otherwise, the four fictitious citation distributions are not unrealistic, in particular, the large number of uncited papers is not uncommon in the usually strongly skewed citation distributions, so that the lowest unique citation count is expected to be zero in most realistic cases. On the other hand, in my example the medium-cited paper, i.e. the 50$^{th}$ paper in datasets A, B, C, D corresponds to the citation counts 1, 0, 2, and 2, respectively, reflecting the different citation levels appertaining to the different fields.

## 4. The first empirical example: a reference set from the Web of Science and the effect of one more and one less citation

In order to calculate the value of the P100 indicator for a specific paper one should obtain the citation distribution of all papers in the same subject category from the same publication year and then determine the position of the paper in this distribution. As an example I shall discuss the citation distribution for articles from 2009 in the subject category "Information Science & Library Science" in the ISI Web of Science. This reference set comprises 3007 papers with 14108 citations (until the end of 2013), yielding 58 unique citation counts, i.e. $i_{max} = 57$. This means the values $100 * i / 57 = 1.7544$ $i$ for the P100 indicator, as shown in Table 3.

There are 15 papers in this set that have been (co-)authored by one of the protagonists of the P100 indicator (L. Leydesdorff). Most of these are (relatively) highly cited and belong to the 10% most cited papers in this reference set, but the P100 values do not reflect this. That is due to the strongly skewed citation distribution: in the present example about 75% of all papers are cited 5 times or less, so that they are condensed into 6 unique citation counts, i.e. about 10% of all unique citation counts. This effect was already discussed by Bornmann et al. (2013). Nevertheless, on average the 15 papers by Leydesdorff obtain a value $\overline{P100}$ = 41.05.

Let us now assume that one of the papers with 40 citations had received **one citation less**. Thus one more unique citation count occurs, leading to changes in all (except the highest and lowest) values of P100 shown as the first modification in Table 3. For Leydesdorff's papers this reduces the average to $\overline{P100}$ = 40.80. Thus a small alteration of one out of 3007 citations in the reference set leads to a much larger change in the performance evaluation of the specific author. And although the reference set is *weakened* by one citation (0.033%), the performance of the author appears to change 18 times more *to the worse* (0.6%). Intuitively I would have expected it to be better than in the original situation.

A similar effect would be obtained, if one of the papers with 44 citations would be **cited once more**: A 0.033% alteration in the reference set then yields a change of the performance result by 1.1% to an average of $\overline{P100}$ = 40.57. But at least in this case the tendency is as expected: An albeit small *improvement* in the reference set yields a *deterioration* of the performance evaluation for the specific author.

Finally, executing both changes (**one citation more and one less**) so that the total number of citations in the reference set does *not change* at all, leads to an even *worse* result $\overline{P100}$ = 40.34, a drop of 1.7% from the original value. This case is shown as the second modification in Table 3.

The discussed effects were caused by filling gaps in the sequence of unique citation counts. In order to improve the performance evaluation result for Leydesdorff one has to execute the opposite, namely deleting unique citation counts. As an example, for the third modification in Table 3 I have assumed that the paper with 67 citations is **cited once more**. Thus one less unique citation count occurs, again all (but the two extreme) values of P100 change and for Leydesdorff's paper the average becomes $\overline{P100}$ = 42.30. As above a small alteration of one citation (0.033% of 3007 citations) in the reference set has a large effect for the author: his evaluation result *improves* 45 times more, namely by 1.50%. This also is in contrast to my expectations: I would rather expect a worsening as a consequence of the albeit small *improvement* in the reference set.

Exactly the same improvement for Leydesdorff can be produced by a small worsening in the reference set assuming for example that the paper with 52 citations had received **one citation less**. This is in agreement with my expectations: a *worse* reference set corresponds to a *better* result for the specific author.

And finally executing both changes (**one citation more and one less**) as shown as the fourth modification in Table 3 yields an average of $\overline{P100}$ = 42.30, an *increase* of 3.05% from the original value, although the total number of citations in the reference set remains *the same*.

### 5. The P100 values for three protagonists of the P100 indicator

In the following I discuss the values of the P100 indicator for the articles of the three senior authors who have proposed the P100 indicator. The results are visualized in Fig. 1 where the unique citation counts for all papers of these authors are utilized and the P100 values for all the papers are shown. Thus especially for low citation counts several values occur for the same citation frequency. The values reflect the position of each publication in the citation distribution of the corresponding reference set which comprises all papers published in the same year and in the same subject category as discussed in the previous section. All citations from the publication year until the end of 2012 have been taken into account, but 30 publications from 2012 are not included in Fig. 1, because the citation window of less than one year is so small that a reliable impact measurement is not possible. If a paper is classified as belonging to more than one subject category, then the P100 value is averaged over the results for the different categories. For example, only 4 of the 15 papers of Leydesdorff which were evaluated in the previous section are attributed solely to the subject category "Information Science & Library Science". The others belong to two subject categories, like most of his articles.

Altogether there are 244 papers with 4581 citations and 55 unique citation counts contributing to Fig. 1. Further 26 uncited papers do not show up in Fig. 1, because the logarithmic scale does not include the value P100 = 0.

Many papers by these authors have a high P100 value reflecting a large impact. It is not surprising that high citation counts usually lead to high values of the P100 indicator. However, the opposite is not always the case: especially for recent publication years relatively low citation counts can correspond to a relatively large impact in comparison with the reference set.

Of course it is not unexpected that older papers with few citations usually end up with small P100 values. But there are exceptions like the two upmost dots in the left-most column in Fig. 1: In these cases a single citation is sufficient to reach a P100 value of 50 or 33.3, because only one or two higher citation frequencies occur in the reference sets, although the papers have been published in 1988 and 1994, respectively. Another noteworthy example is the highest data point for 16 citations in Fig. 1: These 16 citations since 1989 are already the maximum in the respective reference set, so that P100 = 100 is obtained.

On the other hand there is a paper from 2007 with 18 citations that ends up below the main body of data points in Fig. 1, because it is attributed to a different subject category, where obviously higher citation frequencies are more common, so that 18 citations in the relatively short period since 2007 lead to a result of P100 = 6.67 only. Likewise, one paper from 2008 with already 27 citations has a value of P100 = 10 only. Both papers were published in chemistry journals and were attributed to the subject category "Chemistry Multidisciplinary". Another paper from 2005 with 7 citations and P100 = 1.34 also falls below the main regime of data points; it belongs to the same subject category, namely "Multidisciplinary Sciences", as a paper from 1997, for which 4 citations mean P100 = 0.52 only, see Fig. 1. And for another paper as many as 45 citations since 2000 also yield a low value of P100 = 33.73, while 41 citations since 1998 can be enough to get to the top P100 = 100, see Fig. 1, too. These examples demonstrate how crucial the attributed subject categories and thus the corresponding reference sets can be. Choosing a particular journal can therefore strongly influence the impact measurement by the P100 indicator, but also in terms of other indicators that are based on percentiles.

Let us consider a paper on citations in chemistry like the above example with 18 citations. Let us make the assumption that the absolute impact in terms of citation frequency would be the same irrespective of whether it is published in a chemistry journal or a scientometrics journal. Now it seems a better strategy to choose the latter, because this will probably fall into a less cited subject category. Consequently, the relative impact measured by comparison with the reference set will be higher than for the chemistry journal.

## 6. The second empirical example: the three scientists and the effect of small changes

Unfortunately, I have no chance to modify all the reference sets utilized in the previous section. Therefore I cannot discuss the influence of small changes in the citation counts on the P100 results for those empirical citation data. As an alternative I have constructed a rather small reference set by combining the citation data for the three investigated authors. This is certainly not the standard practice in bibliometrics, but in my view it does also make some sense. For example, if one wants to evaluate a scientist in comparison with all colleagues in an institute, it is not unreasonable to use the citation data of the institute as a reference set. In this way one can avoid the problem how to obtain the large number of different citation distributions of all papers published in the same year and in the same subject category as one of the papers in the data set under evaluation. Restricting oneself to the smaller reference set, the data acquisition is simple and inexpensive.

For the following analysis I have downloaded the data from the ISI Web of Knowledge on 19 July 2013. The results are given in Table 4. Altogether there are 365 papers with 5833 citations and 64 unique citation counts which are considered as the reference set for the subsequent discussion. Even restricting the data to citations until the end of 2012 and publications until the end of 2011 as for Fig. 1, the remaining 301 papers with 5261 citations and 60 unique citation counts are considerably more than and more often cited than those presented in Fig. 1. One reason is that I have not restricted my search to articles but included other papers like conference proceedings. But this does not explain the very different citation frequencies. As I cannot influence the search parameters for the data used in Fig. 1, I cannot imagine why so many citations are missing in Fig. 1. For example, the 7 most cited papers in both datasets are the same, but the citation counts in Table 4 (corrected for citations in 2013) are larger by 182, 19, 3, 7, 16, 3, 14 citations, respectively, in comparison with the values in Fig. 1.

In the following, the larger values are used. The respective distributions are visualized in Fig. 2 where all citation counts between 0 and 100 are utilized, and in Fig. 3 where only the 65 citation counts with a non-zero paper count are included. Therefore the curve cannot drop below 1 in this plot. The above mentioned reduction of the number of citation counts can be clearly seen by comparing Figs. 2 and 3. Out of the 101 data between 0 and 100 citations in Fig. 2 only 57 unique citation counts remain in Fig. 3 with 8 more data covering the extended range up to 638 citations.

Let us now assume a few small changes in the citation data of the second author, namely that the number of 134 citations to the third paper is reduced by two, that to the seventh paper is increased from 73 to 74, the two papers with 60 and 62 citations are given 61 citations each, the two papers with 49 and 47 citations are given 48 citations each, the two papers with 39 and 41 citations are given 40 citations each, the papers with 36 and 31 citations are given one more citation, and the paper with 35 citations one less. As a consequence the number of unique citation frequencies is lowered from 65 by 10 to 55. Note that I have chosen the changes in such a way that the total number of citations does not change. However, all the P100 values (except 0 and 100) change, in particular for the smaller citation counts below 60 the P100 values increase which in contrast to my expectations means a better performance evaluation for all authors. For the case of 30 citations the P100 indicator increases by as much as 8.4 percentage points. In conclusion, a small rearrangement of 12 out of 4339 citations for this author produces a remarkable adjustment of the values of the P100 indicator.

For this modification I have hypothesized small fluctuations which altogether did not change the total number of citations. One could achieve a similar effect (although in the present empirical dataset not quite as large), if one would *only increase* citation counts for some selected highly cited papers (not shown). This could be interpreted as reflecting the citation distribution at some later point in time. Thus the slightly larger number of citations for the second author would mean an improved performance evaluation result for all three authors. For me, this appears to be counterintuitive: the performances of the first and the third author did not change, but the P100 indicator values increase and thus lead to an improved result of an evaluation based on these values.

But the inverse effect can also be produced. For this purpose let us again exercise a few small changes in the citation distribution of the second author. In particular, let us take away one citation from one of the papers with 68 citations, from one of those with 58 citations, and from one of those with 27 citations, and let us further add one and two citations to the two papers with originally 44 citations. Again the total number of citations remains unchanged. The results are shown as the second modification in Table 4. Now five more unique citation counts occur and consequently all the P100 values (except 0) below 58 citations are somewhat reduced. For 25 citations the change is as big as 2.9 percentage points. This means again in contrast to my expectations that the overall performance of all three authors appears to be worse than in the original situation. Once more, a small rearrangement of 6 citations leads to a sizable adjustment of P100.

A nearly identical result could be obtained, if one would *only add* citations to the mentioned papers (except for the paper with 27 citations for which instead the paper with 25 citations has to get an additional citation). But then the resulting performance worsening of the first and third author might be considered to be realistic, because the citation performance of the second author has improved and thus the relative performance has improved. However, it is counterintuitive to me that also the overall performance of the second author seems to be worse than in the original situation, as indicated by the lower P100 values for nearly all of his papers.

In Fig. 4 the distribution in the original situation and the modifications from Table 4 are visualized by plotting the numbers of papers versus the percentiles as given by P100. The different number of unique citation counts leads to a slightly wider or narrower spacing of the P100 values for the two modifications in comparison with the original situation. Correspondingly most of the data points for the two modifications are shifted to higher or lower values of P100, respectively. This reflects the discussed better or worse performance evaluation results, respectively, because nearly all of the papers of the three authors are affected.

## 7. The third empirical example: Comparing different publication years for a physics journal

I have previously analyzed the citation distributions of the physics journal Europhysics Letters/EPL for 25 different publication years (Schreiber, 2013c). The data have been retrieved from the ISI Web of Knowledge on 18 July 2012. For the present purpose I consider those data, but only for 4 different publication years. Selected data are presented in Table 5. In particular for each year the five lowest and highest citation counts are given together with their ranks and the resulting values of the P100 indicator. The corresponding paper counts are also shown and the cumulated paper counts (i.e. the sum of the number of papers up to a certain citation count) yields the cumulated percentage, i.e. the percentage of papers up to and including the respective citation count.

Of course the number of lowly cited papers is significantly lower in the earlier years, because these papers had much more time to acquire citations. For the number of unique citation values there is no clear trend observable, as can be seen in Table 5 where the maximal rank increases for 80 in 1986 to 98 in 1987, and likewise from 43 in 2007 to 56 in 2008. One would expect that the highest citation count is larger for the earlier years and that the unique citation counts are spread more widely. Overall this trend can be observed, but in some cases it is counterbalanced by an increase in the total number of papers. The large differences between the top citation counts and the respective top ranks indicate that many citation counts do not occur, because there are no papers with such numbers of citations.

As in Section 3 my main point of interest is the investigation of the top-10% papers in terms of numbers of citations. This corresponds to a cumulated percentage of 90% of the papers. The respective thresholds in the citation counts are included in Table 5. In 1986 for example 65 citations are necessary to reach the top 10%. However, in terms of the unique citation counts this means rank 59 out of 80, which results in a P100 value of 73.75. Accidentally, the 90% threshold for 1987 is given by the same citation frequency. But now in terms of the unique citation counts this number corresponds to rank 61 out of 98, so that the P100 indicator yields a much lower value, namely 62.2. An author of a publication with 65 citations would just make it into the top 10% in both years, but would have obtained a much lower ranking by P100 in 1987 than in 1986. In my opinion, this is unfair.

An even more striking example is obtained by comparing the situation in 2007 and 2008. In 2007 the 90% threshold is determined by 21 citations corresponding to rank 21 out of 43, so that P100 = 48.8 is approximately given by the median. In 2008 the threshold is already reached with 16 citations, corresponding to rank 16 out of 56 so that we get a value of 28.6 for P100. Again an author who is proud of having just made into the top-10% highly cited papers in 2008 would be very disappointed by the performance evaluation in terms of P100 because here the performance seems to be much worse than that of a competitor who just made it into the top 10% in 2007.

## 8. Conclusions

The presented examples show that care must be taken when using the new P100 indicator for performance evaluations. Some problems arise because gaps occur in the sequence of unique citation counts due to the fact that especially for higher citation frequencies more and more often no papers with the respective number of citations exist in the dataset. Then small fluctuations can easily influence the total number of unique citation counts and thus the P100 values. This was already mentioned as a reliability problem by Bornmann and Mutz (2014).

Another problem occurs when different reference sets are used. Such an approach is highly recommendable when one wants to compare different fields and/or different publication years. But the mentioned strong reduction of the data by the usage of unique citation counts means that very different reference sets can have the same or nearly the same values of the P100 indicator. This makes a comparison of researchers by P100 in different fields or for different publication years very difficult if not impossible. A meaningful comparison is certainly possible for researchers contributing to the same reference set. However, also in this case small changes in the high-citation-frequency tail of the distribution can yield surprising effects, leading to counterintuitive behavior when the time evolution of the indicators for different scientists is compared.

In conclusion, the new indicator has clear advantages, but also so strong disadvantages that it might be difficult to apply it in performance evaluations. More empirical investigations are necessary to determine how valuable the new indicator can be.

Bornmann and Mutz (2014) have already shown with an example that P100 can behave paradoxically, and they have therefore introduced the P100' indicator. Already in the original proposal of P100 (Bornmann, Leydesdorff, and Wang, 2013) in which it was shown that P100 behaves very different from usual percentile ranks it was concluded (p. 943) that "P100 underperformes empirically although it is conceptionally convincing." This has been corroborated by the present investigation. The question, whether the refinement to P100' is an improvement and yields a better empirical performance needs to be answered in further empirical studies.

**Acknowledgement**

I am grateful to Lutz Bornmann for providing the values of the P100 indicator, which were utilized in Sections 4 and 5. Those data have been determined from an MPG data base that allows one to add certain indicators to the Web-of-Science records.8

**References**

**Bornmann, L. (2012).** Redundancies in H index variants and the proposal of the number of top-cited papers as an attractive indicator. Measurement, 10, 149-153.

**Bornmann, L. & Mutz, R. (2014).** From P100 to P100': Conception and improvement of a new citation-rank approach in bibliometrics. Accepted for publication in Journal of the American Society for Information Science and Technology, arXiv:1307.0667v2

**Bornmann, L., Leydesdorff, L., & Mutz, R. (2013)**. The use of percentiles and percentile rank classes in the analysis of bibliometric data: opportunities and limits. Journal of Informetrics 7(1), 158-165.

**Bornmann, L., Leydesdorff, L., & Wang, J. (2013).** Which percentile-based approach should be preferred for calculating normalized citation impact values? An empirical comparison of five approaches including a newly developed one (P100). Journal of Informetrics, 7(4), 933-944.

**Hazen, A. (1914).** Storage to be provided in impounding reservoirs for municipal water supply. Transactions of American Society of Civil Engineers, 77, 1539-1640.

**Hirsch, J.E. (2005).** An index to quantify an individual's scientific research output. Proceedings of the National Academy of Sciences, 102(46), 16569-16572.

**Hirsch, J.E. (2007).** Does the h-index have predictive power? Proceedings of the National Academy of Sciences, 104(49), 19193-19198.

**Lehmann, S., Jackson, A.D., & Lautrup, B.E. (2008).** A quantitative analysis of indicators of scientific performance. Scientometrics, 76(2), 369-390.

**Leeuwen, T. (2008).** Testing the validity of the Hirsch-index for research assessment purposes. Research Evaluation, 17(2), 157-160.

**Schreiber, M. (2013a).** Uncertainties and ambiguities in percentiles and how to avoid them. Journal of the American Society for Information Science and Technology, 64(3), 640-643.

**Schreiber, M. (2013b).** Empirical evidence for the relevance of fractional scoring in the calculation of percentile rank scores. Journal of the American Society for Information Science and Technology, 64(4), 861-867.

**Schreiber, M. (2013c).** How much do different ways of calculating percentiles influence the derived performance indicators? - A case study. Scientometrics 97, 821-829.

**Table 1.** Unique citation counts, paper counts, resulting ranks, and corresponding P100 values for the original dataset taken from Bornmann and Mutz (2014) but without the uncited paper in that set; and two modifications as described in the text. If the paper count is zero, then the citation count is not relevant and therefore then there is no rank and no P100 value given in the table.

| Unique citation counts | Original data set | | | First modification | | | Second modification | | | | | |
|---|---|---|---|---|---|---|---|---|---|---|---|---|
| | Paper counts | Rank $i$ | P100 | Paper counts | Rank $i$ | P100 | Paper counts | Rank $i$ | P100 | X | Y | Z |
| 0  | 0 | -  | -   | 0 | -  | -   | 0 | -  | -   | 0 | 0 | 0 |
| 1  | 1 | 0  | 0   | 0 | -  | -   | 1 | 0  | 0   | 1 | 0 | 0 |
| 2  | 1 | 1  | 20  | 2 | 0  | 0   | 1 | 1  | 17  | 0 | 1 | 0 |
| 3  | 1 | 2  | 40  | 1 | 1  | 25  | 1 | 2  | 33  | 0 | 1 | 0 |
| 4  | 3 | 3  | 60  | 3 | 2  | 50  | 2 | 3  | 50  | 0 | 2 | 0 |
| 5  | 0 | -  | -   | 0 | -  | -   | 1 | 4  | 67  | 0 | 1 | 0 |
| 7  | 1 | 4  | 80  | 1 | 3  | 75  | 1 | 5  | 83  | 0 | 0 | 1 |
| 10 | 1 | 5  | 100 | 1 | 4  | 100 | 1 | 6  | 100 | 1 | 0 | 0 |

**Table 2.** Citation counts, resulting ranks, and corresponding P100 values for 4 fictitious datasets which differ by the paper counts as denoted by A, B, C, D. The top-10% most cited papers in each case are indicated by bold face.

| Citation counts | Rank $i$ | P100 | Paper counts | | | |
|---|---|---|---|---|---|---|
| | | | A | B | C | D |
| 0  | 0  | 0   | 45 | 60 | 25 | 26 |
| 1  | 1  | 10  | 20 | 20 | 20 | 20 |
| 2  | 2  | 20  | 10 | 5  | 15 | 20 |
| 3  | 3  | 30  | 7  | 4  | 11 | 10 |
| 4  | 4  | 40  | 5  | **3** | 8 | 7 |
| 5  | 5  | 50  | **4** | **2** | 6 | 5 |
| 6  | 6  | 60  | **3** | **2** | 5 | **4** |
| 7  | 7  | 70  | **2** | **1** | **3** | **3** |
| 8  | 8  | 80  | **2** | **1** | **3** | **2** |
| 9  | 9  | 90  | **1** | **1** | **2** | **2** |
| 10 | 10 | 100 | **1** | **1** | **2** | **1** |

**Table 3.** Unique citation counts and corresponding paper counts for articles in the subject category "Information Science & Library Science" published in 2009. The corresponding rank yields the P100 value. If the paper count is zero, then the citation count is not relevant and therefore then there is no rank and no P100 value given in the table. Data lines to which Leydesdorff contributed by one of his papers (or by two papers in the case of 7 citations) are set in italics. Four modifications in the paper counts are given with the resulting ranks and P100 values. Altered values are indicated by bold face. The citation counts between 28 and 37 are comprised in one line in order to shorten the table.

| No. cits. | Original data set | | | First modification | | | Second modification | | | Third modification | | | Fourth modification | | |
|---|---|---|---|---|---|---|---|---|---|---|---|---|---|---|---|
| | No. paps. | rank | P100 | No. paps. | rank | P100 | No. paps. | rank | P100 | No. paps. | rank | P100 | No. paps. | rank | P100 |
| 0 | 834 | 0 | 0.0 | 834 | 0 | 0.0 | 834 | 0 | 0.0 | 834 | 0 | 0.0 | 834 | 0 | 0.0 |
| 1 | 506 | 1 | 1.8 | 506 | 1 | 1.7 | 506 | 1 | 1.7 | 506 | 1 | 1.8 | 506 | 1 | 1.8 |
| 2 | *353* | *2* | *3.5* | *353* | *2* | *3.4* | *353* | *2* | *3.4* | *353* | *2* | *3.6* | *353* | *2* | *3.6* |
| 3 | 222 | 3 | 5.3 | 222 | 3 | 5.2 | 222 | 3 | 5.1 | 222 | 3 | 5.4 | 222 | 3 | 5.5 |

| cit | N | rank | P100 | N | rank | P100 | N | rank | P100 | N | rank | P100 | N | rank | P100 |
|---|---|---|---|---|---|---|---|---|---|---|---|---|---|---|---|
| 4 | 179 | 4 | 7.0 | 179 | 4 | 6.9 | 179 | 4 | 6.8 | 179 | 4 | 7.1 | 179 | 4 | 7.3 |
| 5 | 153 | 5 | 8.8 | 153 | 5 | 8.6 | 153 | 5 | 8.5 | 153 | 5 | 8.9 | 153 | 5 | 9.1 |
| 6 | 83 | 6 | 10.5 | 83 | 6 | 10.3 | 83 | 6 | 10.2 | 83 | 6 | 10.7 | 83 | 6 | 10.9 |
| 7 | 92 | 7 | 12.3 | 92 | 7 | 12.1 | 92 | 7 | 11.9 | 92 | 7 | 12.5 | 92 | 7 | 12.7 |
| 8 | 65 | 8 | 14.0 | 65 | 8 | 13.8 | 65 | 8 | 13.6 | 65 | 8 | 14.3 | 65 | 8 | 14.5 |
| 9 | 57 | 9 | 15.8 | 57 | 9 | 15.5 | 57 | 9 | 15.3 | 57 | 9 | 16.1 | 57 | 9 | 16.4 |
| 10 | 62 | 10 | 17.5 | 62 | 10 | 17.2 | 62 | 10 | 16.9 | 62 | 10 | 17.9 | 62 | 10 | 18.2 |
| 11 | 60 | 11 | 19.3 | 60 | 11 | 19.0 | 60 | 11 | 18.6 | 60 | 11 | 19.6 | 60 | 11 | 20.0 |
| 12 | 37 | 12 | 21.1 | 37 | 12 | 20.7 | 37 | 12 | 20.3 | 37 | 12 | 21.4 | 37 | 12 | 21.8 |
| 13 | 44 | 13 | 22.8 | 44 | 13 | 22.4 | 44 | 13 | 22.0 | 44 | 13 | 23.2 | 44 | 13 | 23.6 |
| 14 | 32 | 14 | 24.6 | 32 | 14 | 24.1 | 32 | 14 | 23.7 | 32 | 14 | 25.0 | 32 | 14 | 25.5 |
| 15 | 21 | 15 | 26.3 | 21 | 15 | 25.9 | 21 | 15 | 25.4 | 21 | 15 | 26.8 | 21 | 15 | 27.3 |
| 16 | 17 | 16 | 28.1 | 17 | 16 | 27.6 | 17 | 16 | 27.1 | 17 | 16 | 28.6 | 17 | 16 | 29.1 |
| 17 | 28 | 17 | 29.8 | 28 | 17 | 29.3 | 28 | 17 | 28.8 | 28 | 17 | 30.4 | 28 | 17 | 30.9 |
| 18 | 20 | 18 | 31.6 | 20 | 18 | 31.0 | 20 | 18 | 30.5 | 20 | 18 | 32.1 | 20 | 18 | 32.7 |
| 19 | 10 | 19 | 33.3 | 10 | 19 | 32.8 | 10 | 19 | 32.2 | 10 | 19 | 33.9 | 10 | 19 | 34.5 |
| 20 | 8 | 20 | 35.1 | 8 | 20 | 34.5 | 8 | 20 | 33.9 | 8 | 20 | 35.7 | 8 | 20 | 36.4 |
| 21 | 15 | 21 | 36.8 | 15 | 21 | 36.2 | 15 | 21 | 35.6 | 15 | 21 | 37.5 | 15 | 21 | 38.2 |
| 22 | 13 | 22 | 38.6 | 13 | 22 | 37.9 | 13 | 22 | 37.3 | 13 | 22 | 39.3 | 13 | 22 | 40.0 |
| 23 | 7 | 23 | 40.4 | 7 | 23 | 39.7 | 7 | 23 | 39.0 | 7 | 23 | 41.1 | 7 | 23 | 41.8 |
| 24 | 12 | 24 | 42.1 | 12 | 24 | 41.4 | 12 | 24 | 40.7 | 12 | 24 | 42.9 | 12 | 24 | 43.6 |
| 25 | 6 | 25 | 43.9 | 6 | 25 | 43.1 | 6 | 25 | 42.4 | 6 | 25 | 44.6 | 6 | 25 | 45.5 |
| 26 | 4 | 26 | 45.6 | 4 | 26 | 44.8 | 4 | 26 | 44.1 | 4 | 26 | 46.4 | 4 | 26 | 47.3 |
| 27 | 4 | 27 | 47.4 | 4 | 27 | 46.6 | 4 | 27 | 45.8 | 4 | 27 | 48.2 | 4 | 27 | 49.1 |
| 28-37 | 36 | ... | ... | 36 | ... | ... | 36 | ... | ... | 36 | ... | ... | 36 | ... | ... |
| 38 | 3 | 38 | 66.7 | 3 | 38 | 65.5 | 3 | 38 | 64.4 | 3 | 38 | 67.9 | 3 | 38 | 69.1 |
| 39 | 0 | | | **1** | **39** | **67.2** | **1** | **39** | **66.1** | 0 | | | 0 | | |
| 40 | 3 | 39 | 68.4 | **2** | **40** | **69.0** | **2** | **40** | **67.8** | 3 | 39 | 69.6 | 3 | 39 | 70.9 |
| 41 | 1 | 40 | 70.2 | 1 | 41 | 70.7 | 1 | 41 | 69.5 | 1 | 40 | 71.4 | 1 | 40 | 72.7 |
| 42 | 1 | 41 | 71.9 | 1 | 42 | 72.4 | 1 | 42 | 71.2 | 1 | 41 | 73.2 | 1 | 41 | 74.5 |
| 43 | 1 | 42 | 73.7 | 1 | 43 | 74.1 | 1 | 43 | 72.9 | 1 | 42 | 75.0 | 1 | 42 | 76.4 |
| 44 | 3 | 43 | 75.4 | 3 | 44 | 75.9 | **2** | **44** | **74.6** | 3 | 43 | 76.8 | 3 | 43 | 78.2 |
| 45 | 0 | | | 0 | | | **1** | **45** | **76.3** | 0 | | | 0 | | |
| 46 | 1 | 44 | 77.2 | 1 | 45 | 77.6 | 1 | 46 | 78.0 | 1 | 44 | 78.6 | 1 | 44 | 80.0 |
| 48 | 1 | 45 | 78.9 | 1 | 46 | 79.3 | 1 | 47 | 79.7 | 1 | 45 | 80.4 | 1 | 45 | 81.8 |
| 50 | 1 | 46 | 80.7 | 1 | 47 | 81.0 | 1 | 48 | 81.4 | 1 | 46 | 82.1 | 1 | 46 | 83.6 |
| 51 | 2 | 47 | 82.5 | 2 | 48 | 82.8 | 2 | 49 | 83.1 | 2 | 47 | 83.9 | **3** | **47** | **85.5** |
| 52 | 1 | 48 | 84.2 | 1 | 49 | 84.5 | 1 | 50 | 84.7 | 1 | 48 | 85.7 | **0** | | |
| 55 | 1 | 49 | 86.0 | 1 | 50 | 86.2 | 1 | 51 | 86.4 | **0** | | | **0** | | |
| 56 | 1 | 50 | 87.7 | 1 | 51 | 87.9 | 1 | 52 | 88.1 | **2** | **49** | **87.5** | **2** | **48** | **87.3** |
| 61 | 1 | 51 | 89.5 | 1 | 52 | 89.7 | 1 | 53 | 89.8 | 1 | 50 | 89.3 | 1 | 49 | 89.1 |
| 65 | 1 | 52 | 91.2 | 1 | 53 | 91.4 | 1 | 54 | 91.5 | 1 | 51 | 91.1 | 1 | 50 | 90.9 |
| 67 | 1 | 53 | 93.0 | 1 | 54 | 93.1 | 1 | 55 | 93.2 | 1 | 52 | 92.9 | 1 | 51 | 92.7 |
| 68 | 1 | 54 | 94.7 | 1 | 55 | 94.8 | 1 | 56 | 94.9 | 1 | 53 | 94.6 | 1 | 52 | 94.5 |
| 111 | 1 | 55 | 96.5 | 1 | 56 | 96.6 | 1 | 57 | 96.6 | 1 | 54 | 96.4 | 1 | 53 | 96.4 |
| 119 | 1 | 56 | 98.2 | 1 | 57 | 98.3 | 1 | 58 | 98.3 | 1 | 55 | 98.2 | 1 | 54 | 98.2 |
| 123 | 1 | 57 | 100.0 | 1 | 58 | 100.0 | 1 | 59 | 100.0 | 1 | 56 | 100.0 | 1 | 55 | 100.0 |

**Table 4.** Unique citation counts and corresponding paper counts for the senior authors (LB, LL, RM) who proposed the P100 indicator. The total number of papers for each citation count determines whether this citation count is unique or not relevant. The corresponding rank yields the P100 value. Two modifications (LL1 and LL2) in the paper counts for the second author are given with the resulting ranks and P100 values. Altered values are indicated by bold face. The citation counts between 4 and 23 are comprised in one line in order to shorten the table. The paper counts for the three

authors often do not add up to the total paper count for a given unique citation count, because many papers were co-authored by two or three of the investigated scientists; e.g. the 21, 46, 11 uncited papers do not add up to a total paper count of 78, but only to 72 as shown in the first data line of the table.

| No. cits. | Original data set |  |  |  |  |  | First modification |  |  |  | Second modification |  |  |  |
|---|---|---|---|---|---|---|---|---|---|---|---|---|---|---|
|  | LB | LL | RM | altogether | rank | P100 | LL1 | altogether | rank | P100 | LL2 | altogether | rank | P100 |
| 0    | 21 | 46  | 11 | 72  | 0  | 0.0   | 46  | 72  | 0  | 0.0   | 46  | 72  | 0  | 0.0   |
| 1    | 17 | 15  | 4  | 32  | 1  | 1.6   | 15  | 32  | 1  | 1.9   | 15  | 32  | 1  | 1.4   |
| 2    | 9  | 16  | 6  | 24  | 2  | 3.1   | 16  | 24  | 2  | 3.7   | 16  | 24  | 2  | 2.9   |
| 3    | 4  | 8   | 1  | 12  | 3  | 4.7   | 8   | 12  | 3  | 5.6   | 8   | 12  | 3  | 4.3   |
| 4-23 | 51 | 105 | 14 | 156 | ...| ...   | 105 | 156 | ...| ...   | 105 | 156 | ...| ...   |
| 24   | 3  | 4   | 0  | 6   | 24 | 37.5  | 4   | 7   | 24 | 44.4  | 4   | 7   | 24 | 34.8  |
| 25   | 2  | 2   | 0  | 3   | 25 | 39.1  | 2   | 3   | 25 | 46.3  | 2   | 3   | 25 | 36.2  |
| 26   | 0  | 0   | 0  | 0   |    |       | 0   | 0   |    |       | 1   | 1   | 26 | 37.7  |
| 27   | 0  | 2   | 0  | 2   | 26 | 40.6  | 2   | 2   | 26 | 48.1  | 1   | 1   | 27 | 39.1  |
| 28   | 1  | 1   | 0  | 2   | 27 | 42.2  | 1   | 2   | 27 | 50.0  | 1   | 2   | 28 | 40.6  |
| 29   | 1  | 1   | 0  | 1   | 28 | 43.8  | 1   | 1   | 28 | 51.9  | 1   | 1   | 29 | 42.0  |
| 30   | 1  | 1   | 1  | 2   | 29 | 45.3  | 1   | 2   | 29 | 53.7  | 1   | 2   | 30 | 43.5  |
| 31   | 0  | 1   | 0  | 1   | 30 | 46.9  | 0   | 0   |    |       | 1   | 1   | 31 | 44.9  |
| 32   | 0  | 3   | 0  | 3   | 31 | 48.4  | 4   | 4   | 30 | 55.6  | 3   | 3   | 32 | 46.4  |
| 33   | 2  | 3   | 1  | 5   | 32 | 50.0  | 3   | 5   | 31 | 57.4  | 3   | 5   | 33 | 47.8  |
| 34   | 1  | 2   | 0  | 3   | 33 | 51.6  | 3   | 4   | 32 | 59.3  | 2   | 3   | 34 | 49.3  |
| 35   | 0  | 1   | 0  | 1   | 34 | 53.1  | 0   | 0   |    |       | 1   | 1   | 35 | 50.7  |
| 36   | 0  | 1   | 0  | 1   | 35 | 54.7  | 0   | 0   |    |       | 1   | 1   | 36 | 52.2  |
| 37   | 1  | 1   | 0  | 2   | 36 | 56.3  | 2   | 3   | 33 | 61.1  | 1   | 2   | 37 | 53.6  |
| 38   | 1  | 1   | 0  | 2   | 37 | 57.8  | 1   | 2   | 34 | 63.0  | 1   | 2   | 38 | 55.1  |
| 39   | 0  | 1   | 0  | 1   | 38 | 59.4  | 0   | 0   |    |       | 1   | 1   | 39 | 56.5  |
| 40   | 1  | 2   | 1  | 2   | 39 | 60.9  | 4   | 4   | 35 | 64.8  | 2   | 2   | 40 | 58.0  |
| 41   | 0  | 1   | 0  | 1   | 40 | 62.5  | 0   | 0   |    |       | 1   | 1   | 41 | 59.4  |
| 42   | 0  | 3   | 0  | 3   | 41 | 64.1  | 3   | 3   | 36 | 66.7  | 3   | 3   | 42 | 60.9  |
| 43   | 0  | 2   | 0  | 2   | 42 | 65.6  | 2   | 2   | 37 | 68.5  | 2   | 2   | 43 | 62.3  |
| 44   | 1  | 2   | 0  | 3   | 43 | 67.2  | 2   | 3   | 38 | 70.4  | 0   | 1   | 44 | 63.8  |
| 45   | 0  | 0   | 0  | 0   |    |       | 0   | 0   |    |       | 1   | 1   | 45 | 65.2  |
| 46   | 0  | 0   | 0  | 0   |    |       | 0   | 0   |    |       | 1   | 1   | 46 | 66.7  |
| 47   | 0  | 1   | 0  | 1   | 44 | 68.8  | 0   | 0   |    |       | 1   | 1   | 47 | 68.1  |
| 48   | 0  | 1   | 0  | 1   | 45 | 70.3  | 3   | 3   | 39 | 72.2  | 1   | 1   | 48 | 69.6  |
| 49   | 0  | 1   | 0  | 1   | 46 | 71.9  | 0   | 0   |    |       | 1   | 1   | 49 | 71.0  |
| 54   | 0  | 1   | 0  | 1   | 47 | 73.4  | 1   | 1   | 40 | 74.1  | 1   | 1   | 50 | 72.5  |
| 57   | 0  | 0   | 0  | 0   |    |       | 0   | 0   |    |       | 1   | 1   | 51 | 73.9  |
| 58   | 0  | 2   | 0  | 2   | 48 | 75.0  | 2   | 2   | 41 | 75.9  | 1   | 1   | 52 | 75.4  |
| 60   | 0  | 1   | 0  | 1   | 49 | 76.6  | 0   | 0   |    |       | 1   | 1   | 53 | 76.8  |
| 61   | 0  | 0   | 0  | 0   |    |       | 2   | 2   | 42 | 77.8  | 0   | 0   |    |       |
| 62   | 0  | 1   | 0  | 1   | 50 | 78.1  | 0   | 0   |    |       | 1   | 1   | 54 | 78.3  |
| 67   | 0  | 0   | 0  | 0   |    |       | 0   | 0   |    |       | 1   | 1   | 55 | 79.7  |
| 68   | 0  | 2   | 0  | 2   | 51 | 79.7  | 2   | 2   | 43 | 79.6  | 1   | 1   | 56 | 81.2  |
| 70   | 0  | 1   | 0  | 1   | 52 | 81.3  | 1   | 1   | 44 | 81.5  | 1   | 1   | 57 | 82.6  |
| 73   | 0  | 1   | 0  | 1   | 53 | 82.8  | 0   | 0   |    |       | 1   | 1   | 58 | 84.1  |
| 74   | 1  | 0   | 0  | 1   | 54 | 84.4  | 1   | 2   | 45 | 83.3  | 0   | 1   | 59 | 85.5  |
| 82   | 0  | 1   | 0  | 1   | 55 | 85.9  | 1   | 1   | 46 | 85.2  | 1   | 1   | 60 | 87.0  |
| 99   | 0  | 1   | 0  | 1   | 56 | 87.5  | 1   | 1   | 47 | 87.0  | 1   | 1   | 61 | 88.4  |
| 106  | 1  | 0   | 1  | 1   | 57 | 89.1  | 0   | 1   | 48 | 88.9  | 0   | 1   | 62 | 89.9  |
| 118  | 1  | 0   | 0  | 1   | 58 | 90.6  | 0   | 1   | 49 | 90.7  | 0   | 1   | 63 | 91.3  |
| 123  | 0  | 1   | 0  | 1   | 59 | 92.2  | 1   | 1   | 50 | 92.6  | 1   | 1   | 64 | 92.8  |
| 128  | 1  | 0   | 0  | 1   | 60 | 93.8  | 0   | 1   | 51 | 94.4  | 0   | 1   | 65 | 94.2  |
| 132  | 1  | 0   | 0  | 1   | 61 | 95.3  | 1   | 2   | 52 | 96.3  | 0   | 1   | 66 | 95.7  |
| 134  | 0  | 1   | 0  | 1   | 62 | 96.9  | 0   | 0   |    |       | 1   | 1   | 67 | 97.1  |
| 149  | 0  | 1   | 0  | 1   | 63 | 98.4  | 1   | 1   | 53 | 98.1  | 1   | 1   | 68 | 98.6  |
| 638  | 0  | 1   | 0  | 1   | 64 | 100.0 | 1   | 1   | 54 | 100.0 | 1   | 1   | 69 | 100.0 |

**Table 5.** Selected citation counts, corresponding ranks and resulting P100 values for the citation record of the physics journal Europhysics Letters/EPL in 4 selected years. The paper counts and the cumulated percentage of the number of papers up to and including the respective citation count are also given.

| 1986 (234 papers) | | | | | 1987 (429 papers) | | | | | 2007 (601 papers) | | | | | 2008 (847 papers) | | | | |
|---|---|---|---|---|---|---|---|---|---|---|---|---|---|---|---|---|---|---|---|
| Citation count | Rank | P100 | Paper count | Cumulated percentage | Citation count | Rank | P100 | Paper count | Cumulated percentage | Citation count | Rank | P100 | Paper count | Cumulated percentage | Citation count | Rank | P100 | Paper count | Cumulated percentage |
| 0 | 0 | 0.0 | 9 | 3.8 | 0 | 0 | 0.0 | 15 | 3.5 | 0 | 0 | 0.0 | 38 | 6.3 | 0 | 0 | 0.0 | 83 | 9.8 |
| 1 | 1 | 1.3 | 5 | 6.0 | 1 | 1 | 1.0 | 22 | 8.6 | 1 | 1 | 2.3 | 69 | 17.8 | 1 | 1 | 1.8 | 99 | 21.5 |
| 2 | 2 | 2.5 | 14 | 12.0 | 2 | 2 | 2.0 | 20 | 13.3 | 2 | 2 | 4.7 | 44 | 25.1 | 2 | 2 | 3.6 | 89 | 32.0 |
| 3 | 3 | 3.8 | 6 | 14.5 | 3 | 3 | 3.1 | 17 | 17.2 | 3 | 3 | 7.0 | 56 | 34.4 | 3 | 3 | 5.4 | 91 | 42.7 |
| 4 | 4 | 5.0 | 5 | 16.7 | 4 | 4 | 4.1 | 22 | 22.4 | 4 | 4 | 9.3 | 51 | 42.9 | 4 | 4 | 7.1 | 67 | 50.6 |
| *5-62* | *5-57* | *...* | *170* | *89.3* | *5-62* | *5-59* | *...* | *289* | *89.7* | *5-19* | *5-19* | *...* | *278* | *89.2* | *5-14* | *5-14* | *...* | *315* | *87.8* |
| 63 | 58 | 72.5 | 1 | 89.7 | 63 | 60 | 61.2 | 1 | 90.0 | 20 | 20 | 46.5 | 4 | 89.9 | 15 | 15 | 26.8 | 9 | 88.9 |
| 65 | 59 | 73.8 | 1 | 90.2 | 65 | 61 | 62.2 | 1 | 90.2 | 21 | 21 | 48.8 | 5 | 90.7 | 16 | 16 | 28.6 | 10 | 90.1 |
| *67-170* | *60-75* | *...* | *18* | *97.9* | *67-345* | *62-93* | *...* | *37* | *98.8* | *22-49* | *22-38* | *...* | *50* | *99.0* | *17-297* | *17-51* | *...* | *79* | *99.4* |
| 212 | 76 | 95.0 | 1 | 98.3 | 349 | 94 | 95.9 | 1 | 99.1 | 50 | 39 | 90.7 | 1 | 99.2 | 340 | 52 | 92.9 | 1 | 99.5 |
| 235 | 77 | 96.3 | 1 | 98.7 | 355 | 95 | 96.9 | 1 | 99.3 | 52 | 40 | 93.0 | 2 | 99.5 | 364 | 53 | 94.6 | 1 | 99.6 |
| 275 | 78 | 97.5 | 1 | 99.1 | 490 | 96 | 98.0 | 1 | 99.5 | 66 | 41 | 95.3 | 1 | 99.7 | 365 | 54 | 96.4 | 1 | 99.8 |
| 314 | 79 | 98.8 | 1 | 99.6 | 520 | 97 | 99.0 | 1 | 99.8 | 80 | 42 | 97.7 | 1 | 99.8 | 399 | 55 | 98.2 | 1 | 99.9 |
| 318 | 80 | 100.0 | 1 | 100.0 | 575 | 98 | 100.0 | 1 | 100.0 | 115 | 43 | 100.0 | 1 | 100.0 | 439 | 56 | 100.0 | 1 | 100.0 |

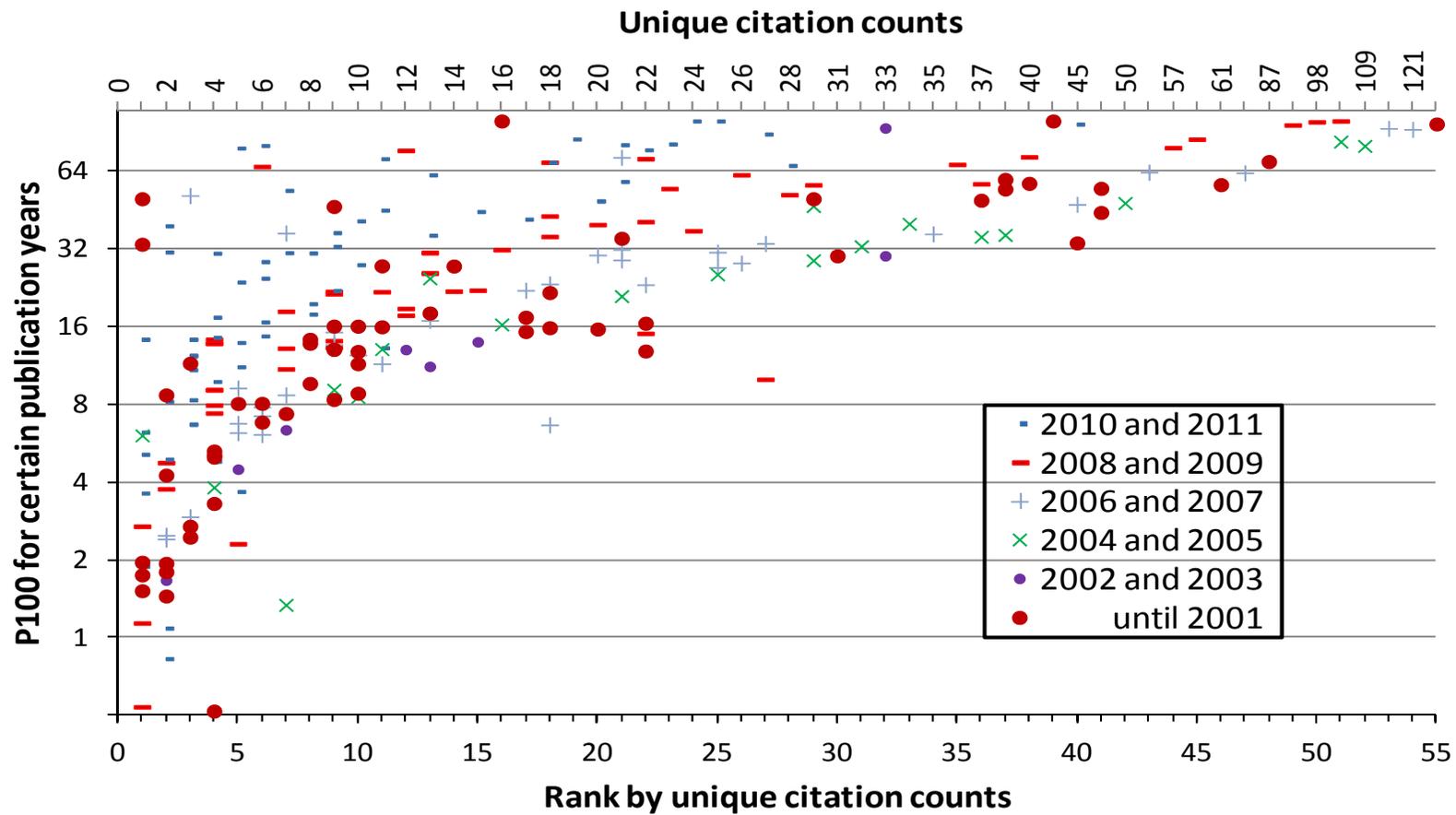

**Fig. 1** Values of the P100 indicator for the articles of the three senior authors. Note the logarithmic scale. Different publication periods are distinguished. Only citation counts for which at least one article was published by one of the authors are utilized.

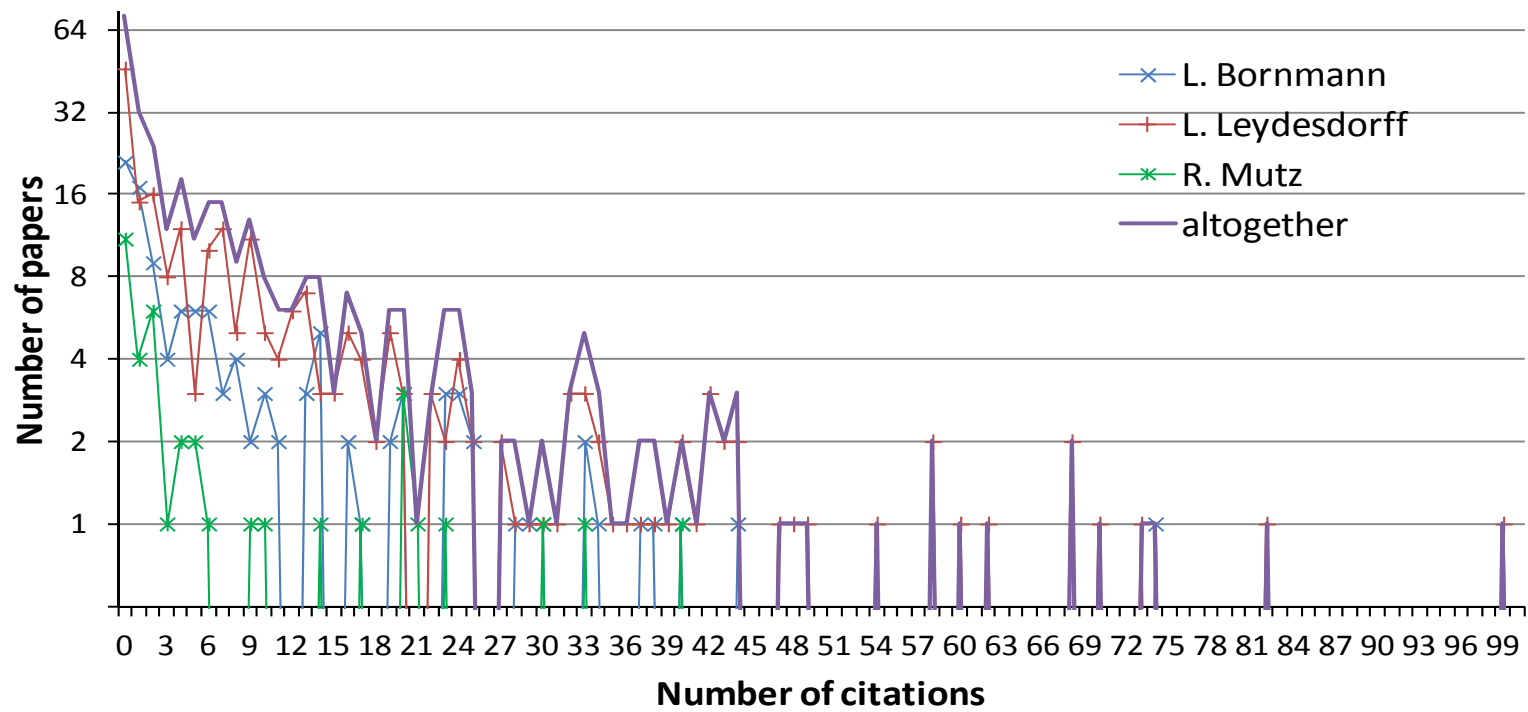

**Fig. 2.** Distribution of papers from the citation record of the three senior authors who proposed the P100 indicator. Note the logarithmic scale. 8 papers with more than 100 citations are not included.

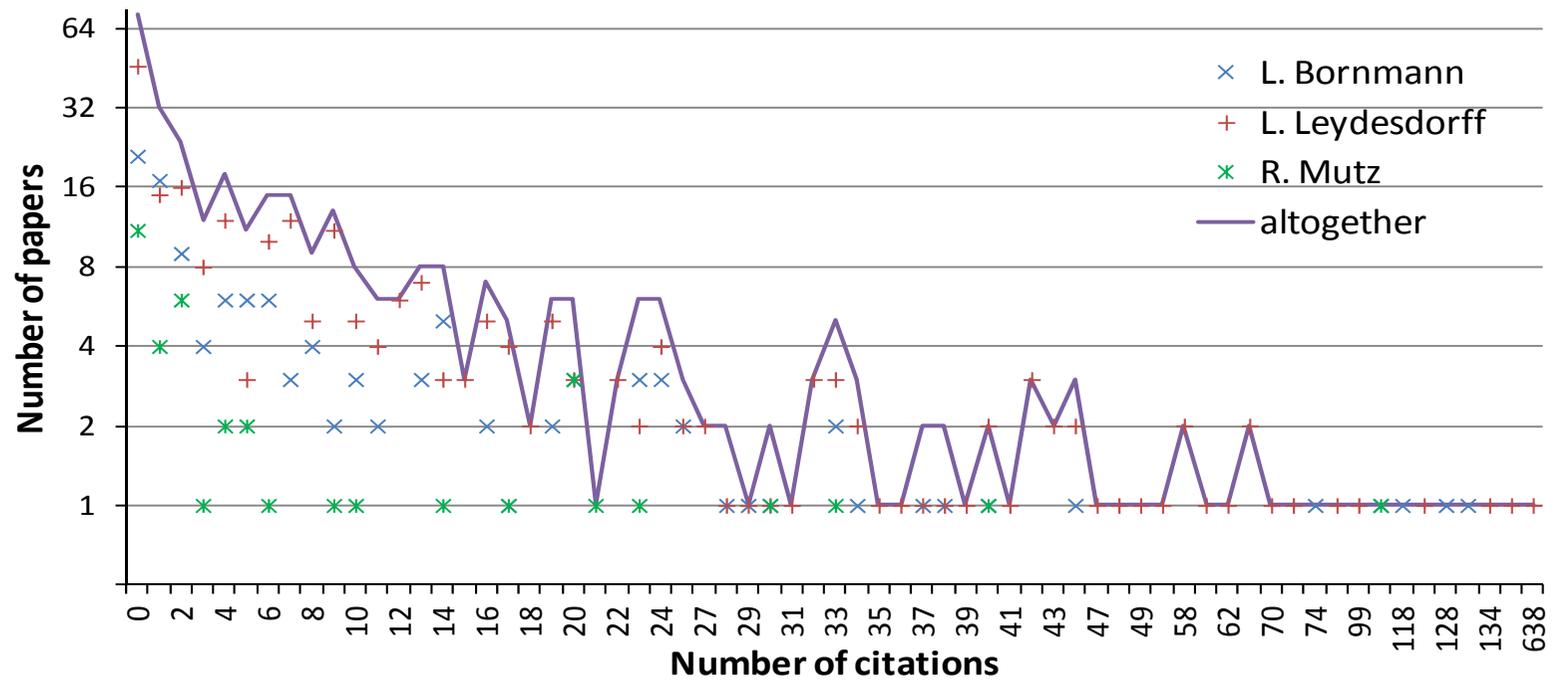

**Fig. 3.** Same as Fig. 2, but only for the unique citation counts.

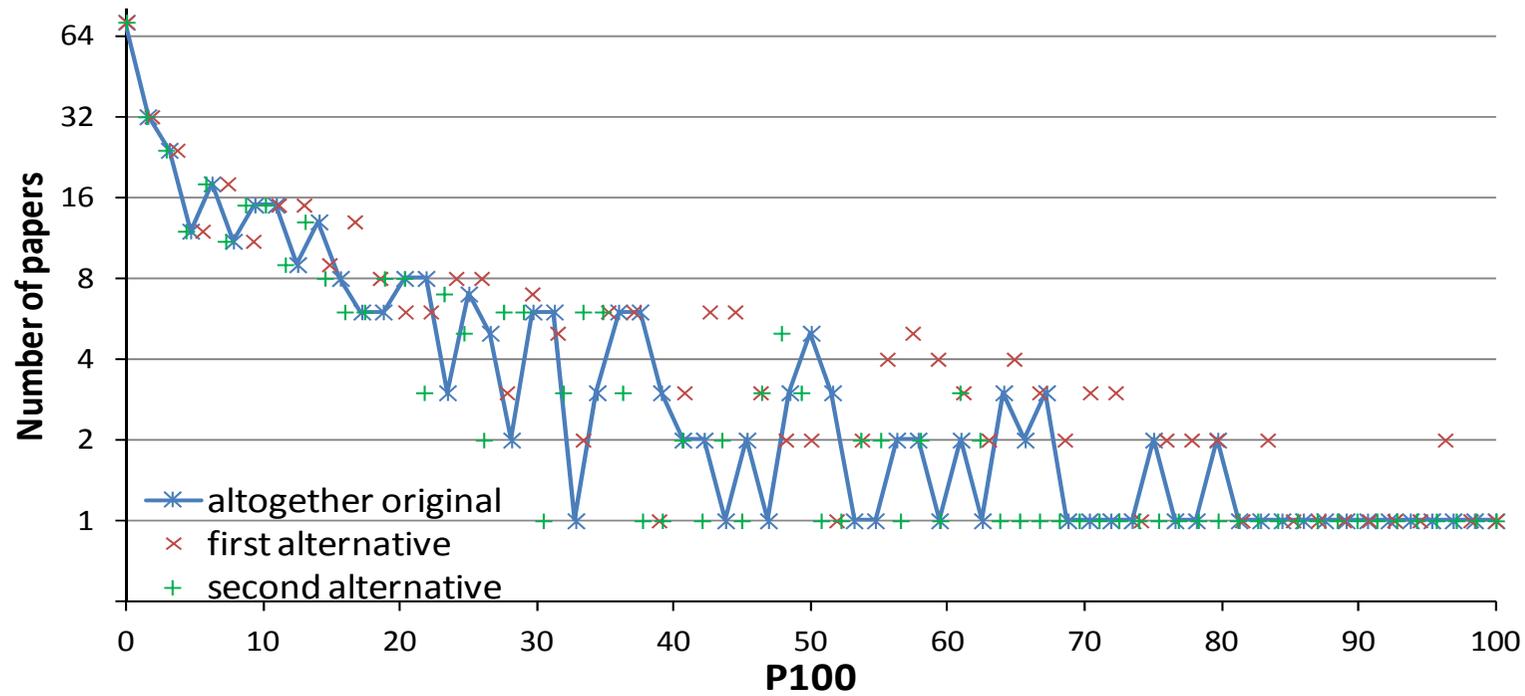

**Fig. 4.** Same as Fig. 3, showing only the sum of the paper counts of the three authors, but also for the two modifications discussed in the text. For the original situation, the P100 scale is equivalent to the citation-frequency scale in Fig. 3.